\documentclass[twocolumn]{elsart}

\usepackage[latin1]{inputenc}
\usepackage{graphicx}
\usepackage{amsmath}
\usepackage{amssymb}
\usepackage{dcolumn}
\usepackage[numbers,sort&compress]{natbib}
\newcolumntype{f}[1]{D{.}{.}{#1}}

\begin{document}
\begin{frontmatter}
\title{Highly charged ion X-rays from Electron-Cyclotron Resonance Ion Sources}

\author[paris]{P.~Indelicato\corauthref{cor}},
\corauth[cor]{Corresponding author}
\ead{paul.indelicato@lkb.ens.fr}
\author[paris]{S. Boucard},
\author[coimbra]{D.S. Covita},
\author[juelich]{D.~Gotta},
\author[wien]{A.~Gruber},
\author[wien]{A.~Hirtl},
\author[wien]{H.~Fuhrmann},
\author[paris]{E.-O. Le Bigot},
\author[paris]{S. Schlesser},
\author[coimbra]{J.M.F. dos  Santos},
\author[psi]{L.~M.~Simons},
\author[psi]{L.~Stingelin},
\author[paris]{M. Trassinelli},
\author[coimbra]{J. Veloso},
\author[psi]{A.~Wasser},
\author[wien]{J.~Zmeskal}

\address[paris]
{Laboratoire Kastler Brossel, École Normale Supérieure et Université
 Pierre et Marie Curie-Paris 6, Case 74, 4 place Jussieu, 
F-75252 Paris, Cedex 05, France}
\address[coimbra]
{Department of Physics, Coimbra University, P-3000 Coimbra, Portugal}

\address[juelich]
{Institut f\"{u}r Kernphysik, Forschungszentrum J\"{u}lich, D-52425 J\"{u}lich,
 Germany}
 \address[wien]
{Institut f\"{u}r Mittelenergiephysik, Austrian Acad. of Sci., Vienna}
\address[psi]
{Paul Scherrer Institut, Villigen PSI, CH5232 Villigen, Switzerland}

\begin{abstract}
Radiation from the highly-charged ions contained in the plasma 
of Electron-Cyclotron Resonance Ion Sources constitutes a very bright source of X-rays.
 Because the ions have a relatively low kinetic energy ($\approx 1$~eV) transitions can be very narrow, containing only
a small Doppler broadening.
  We describe preliminary accurate measurements of
 two and three-electron ions with $Z=16$--18. We show how these measurement can test sensitively many-body
 relativistic calculations or can be used as X-ray standards for precise measurements of X-ray transitions
 in exotic atoms. 

\end{abstract}
\end{frontmatter}


\section{Introduction}
X-rays from X-ray tubes and targets excited by fluorescence have been
 used for a century to provide information on atomic and solid state structure,
 and as wavelength standards. Relativistic correlation effects and quantum-electrodynamics (QED) contributions
 have been the subject of numerous work that have lead after more than a quarter
 of a century of work to the publication of a new X-ray table, containing both all
 experimental values since the 1920's and advanced atomic calculations \citep{dkib2003}.
Yet all this considerable work does not allow for precise-enough tests of
 QED and relativistic effects in high fields, when the speed of the electron
$Z\alpha c$ and the strength of the coupling with the nuclear charge $Z\alpha$ start to get large  ($\alpha$ is the fine structure constant, $c$ is the speed of light and $Z$ the atomic number).
Lamb shift values have been extracted from experimental K$\alpha$ line energies up to $Z=100$  \citep{ibl1998}, 
but neither the theoretical accuracy nor the experimental one allows such data to make very good tests of
 QED in high-field.
 Moreover, because of the complex nature of inner-shell transitions in normal atoms,
 particularly when embedded in a solid matrix, their X-rays are not as suited as
 one would like as X-ray standards. In particular the lines are broad because of
 the Auger effect, deformed because of the outer shell structure and multivacancies,
 and can be shifted because of chemical shifts and of their dependence on the excitation energy.

In the last few years, the need of better X-ray standards has shown up in a series of experiments, performed at
the Paul Scherrer Institute in Switzerland. The aim of these experiments was to make accurate measurements of the
charged pion mass using pionic atoms, and of the strong interaction shift and broadening of the $1s$ level of pionic hydrogen.
Pionic atoms are a specific example of exotic atoms, in which a stable particle (by this we mean a particle that can live long enough to form an atom) is bound to
a nucleus. For light elements at least, the cascade following the capture of the particle, which is always much heavier than the electron, leads to a two body system,
 all the electrons being ejected by Auger effect.

The goal of these experiments was in one hand to provide a pion mass accurate
 to $\approx 2$~ppm, and in the other hand to measure the
strong interaction broadening of the ground state level of 
pionic hydrogen of about 1 eV to an accuracy 
of less than 10 meV, an order of magnitude better than previous experiments \citep{sbgj2001}.
 Such an accuracy is required to obtain a meaningful test of Chiral Perturbation
theory (ChPT) calculations \citep{wei1979,gilm2003}, an effective field theory designed to perform Quantum Chromodynamics (QCD) calculation at low energy, a regime in which quark confinement precludes the use
of perturbation theory in term of the QCD  coupling constant. More recently the interest of doing similar measurements on pionic deuterium was pointed out \citep{mrr2005}
 and the experiment was performed in the summer of 2006.

The previous measurements of the charged pion mass were performed by stopping a pion beam into a solid Mg target\citep{jnbd1986} and measuring the energy
of a transition between two circular, high principal quantum number levels to avoid influence from strong interaction.
The pionic magnesium thus formed was able to recapture electrons from the solid, leading to a difficult analysis and to doubts on the
reliability of the results\citep{dfhj1991,jgl94,abdf96}. A first experiment using  a device called the Cyclotron Trap \cite{sim1993}, designed to slow down beams
of exotic particles and stop them in a nitrogen gas target, was then performed by our collaboration to remove this ambiguity\citep{lbgg1998}.
The K$\alpha$ spectrum of copper, in the fourth order of diffraction
was used as a X-ray standard. This contributed noticeably to the final error budget. It was then decided to improve on this experiment by doing a direct comparison
bewteen the $5g\to 4f$ transitions in muonic oxygen and pionic nitrogen, which are very close in energy. A new cyclotron trap (Cyclotron Trap II) was designed to optimize the 
capture of muons, which comes from the disintegration of the pions in the trap. A demonstration of the
use of exotic atom, and a proposal to use highly-charge ions as X-ray standards was done in Ref.~\cite{agis2003}.

The strong interaction shift ($\approx 7$~eV) and broadening ($\approx 1$~eV) in pionic hydrogen are obtained  by measuring the energy and line shape  of
 X-ray transitions feeding the atomic $1s$ state \citep{abbd2003}. 
The energies of the X-rays in question are 2.436, 
2.886 and   3.043~keV  for the 2p$\to$1s, 3p$\to$1s 
and  4p$\to$1s transitions,
respectively. In order to attain the desired accuracy, a characterization of the spectrometer was in order, using several lines of energy close to those three 
pionic lines, and of  width negligible compared to the best Bragg crystals energy resolution. It was found that the so called relativistic M1 transition
 $1s 2s \,^3S_1\to 1s² \,^1S_0$ (it has exactly 0 probability in a non-relativistic model) of sulfur,
chlorine and argon had the exact energy required. Moreover their natural width of a few neV was completely negligible. It was also found that the 
magnetic field configuration of the improved cyclotron trap was perfect to turn it into a high-performance Electron Cyclotron Resonance ion source (ECRIS).
A single observation of X-ray transitions in helium-like argon had been performed at that time \citep{dkgb2000}, and it was clear from the formation mechanism of
the different charge states that the M1 transition X-ray emission should be very bright in a high-performance source \citep{mcsi2001}. A specific set of polar pieces
and a permanent-magnet hexapole were constructed to enable to turn the cyclotron trap II into an electron-cyclotron resonance ion trap (ECRIT) \citep{bsh2000}. This device
has a very high magnetic field mirror ratio along its axis to improve the electron confinement, leading to very high-performances with a relatively modest frequency for the
microwave (6.4~GHz) driving the electron-cyclotron resonance. The name ECRIT rather than ECRIS stems from the fact that our aim was to improve X-ray emission from the plasma 
inside the source, not to extract intense ion beams. 

Here we provide a preliminary account of  a new measurement of the 
$1s 2s 2p\,^2P_J \to 1s^2 2s\,^2S_{1/2}$ transitions in lithium-like sulfur, chlorine and argon, as an example
of the accuracy that can be reached when combining the ECRIT with a high-performance spectrometer. 

\section{Experimental set-up}

The experiment uses a Bragg spectrometer in Johann mounting \citep{joh1931}
and was equipped  with 
 spherically bent silicon or quartz crystals having a diameter of 
100 mm and a thickness of 0.3 mm or 0.2 mm, respectively. They were 
mounted by optical contact on glass lenses of ultimate quality. The curvature 
radius $R_C$ of the bent crystals is $R_C \approx 2982$~mm and varies between individual crystals. These radii have been recently individually measured to the required accuracy.  
For these parameters,
bent crystal theory predicts a negligible influence of 
the bending process on the crystal's rocking curve \citep{hwf1998}.

 \begin{figure*}[t]
 \centering
 \includegraphics[angle=0,scale=0.5]{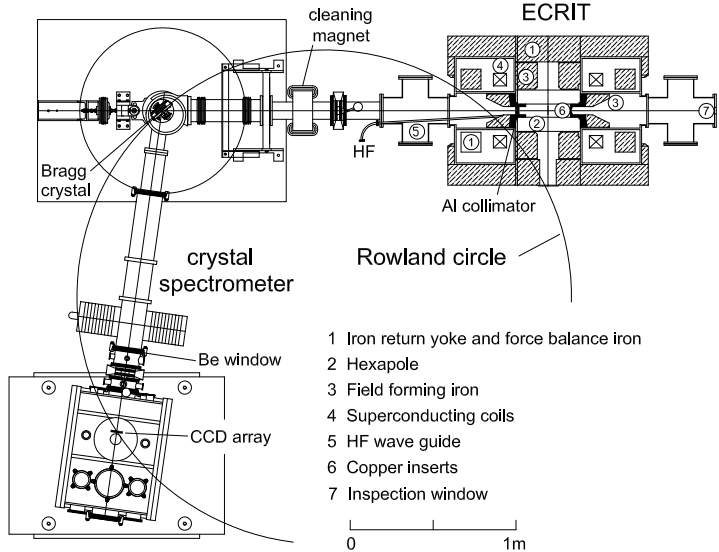}
 \caption[]{
 Set up of the PSI ECRIT together with the Bragg crystal spectrometer 
 }
 \label{fig:ecrit+spect}
 \end{figure*}

The experiment (Fig.~\ref{fig:ecrit+spect}) is composed of
three parts:
\begin{itemize}
\item 
 The Electron Cyclotron Resonance Ion Trap  (ECRIT)  
 consists out of a superconducting 
split coil magnet, which together with special iron inserts, provides the 
mirror field configuration, an Advanced ECR source - Updated version (AECRU-U) 
style permanent  hexapole magnet  and a 6.4~GHz 
power regulated microwave emitter \citep{xie1998}.  The mirror field
 parameters provide 
one of the highest mirror ratios for ECR sources with a value of 4.3 
over the length of the plasma chamber. The plasma chamber is formed by a 0.4 mm thick stainless steel 
tube of inner diameter of 85 mm and a length of 265 mm axially limited 
by copper inserts. At the position of the hexapole gap the stainless 
steel tube is perforated by a series of  2.5 mm diameter holes allowing 
for radial pumping in addition to axial pumping.
The microwave high frequency power 
 is introduced directly to the plasma chamber with waveguides exhibiting 
 a small angle to the axis. In this way the path of the X-rays was at no point 
 cut by any obstacle which could distort the shape of the response function.
 An extraction voltage of 2 kV had been routinely applied at the  side
 opposite to the crystal spectrometer. The total 
ion current was  measured  as a control  for a stable operation. More details 
on the ECRIT can be found in \citep{bsh2000,abgg2005}.

 A reference pressure (without plasma) of 1.7 $10^{-7}$ mbar was achieved. Gas 
filling was supplied radially by UHV precision leak valves  through 
the gaps in  the open structure hexapole.  The gas composition was routinely surveyed with a quadrupole 
mass spectrometer.
For an optimised plasma source a drastic increase of the number
 of energetic electrons was discovered which required 
the use of a cleaning magnet installed at a distance of one meter in 
front of the crystal. 
\item
 A silicon(111) as well as  a quartz(10$\bar{1}$) crystal were  investigated,
 which had been recently applied 
for measuring  pionic hydrogen transitions  
\citep{ghlm2003}.
   The Bragg angles $\Theta_B$, corresponding to  the M1 transition of helium-like argon
 with an energy of 3.104 keV, are $\Theta_B$= 36.68$^\circ$ for 
the quartz and $\Theta_B$=39.57$^\circ$ 
for the silicon crystal.
 The  crystals  were installed at a distance
 of 2330 mm from the centre of the ECRIT resulting in a position of the 
plasma about 500 mm outside the Rowland circle.

\item
 A Charged Coupled Device (CCD)  pixel detector with a pixel 
size of 40~$\mu$m$×$­40~$\mu$m and an energy 
resolution of 140 eV at 3 keV was used to detect the  
X-rays \citep{naab2002}. The detector consisted out of six chips with 600$×$600
pixels each resulting in a total height 
of 72~mm and a width of 48~mm. 
 The distance of the CCD detector from the crystal could be changed remotely
 over a length of 86 mm without breaking the vacuum.

 The CCD  detector  
and the associated electronics were  protected against light as well as 
the high frequency stray field  by a 30 $\mu$m thick beryllium window
 installed in the 
vacuum tube in front of the CCD cryostat.

\end {itemize}

\section{Results and discussion}

\begin{figure}[b]
\centering
\includegraphics[width=4.5cm,angle=270]{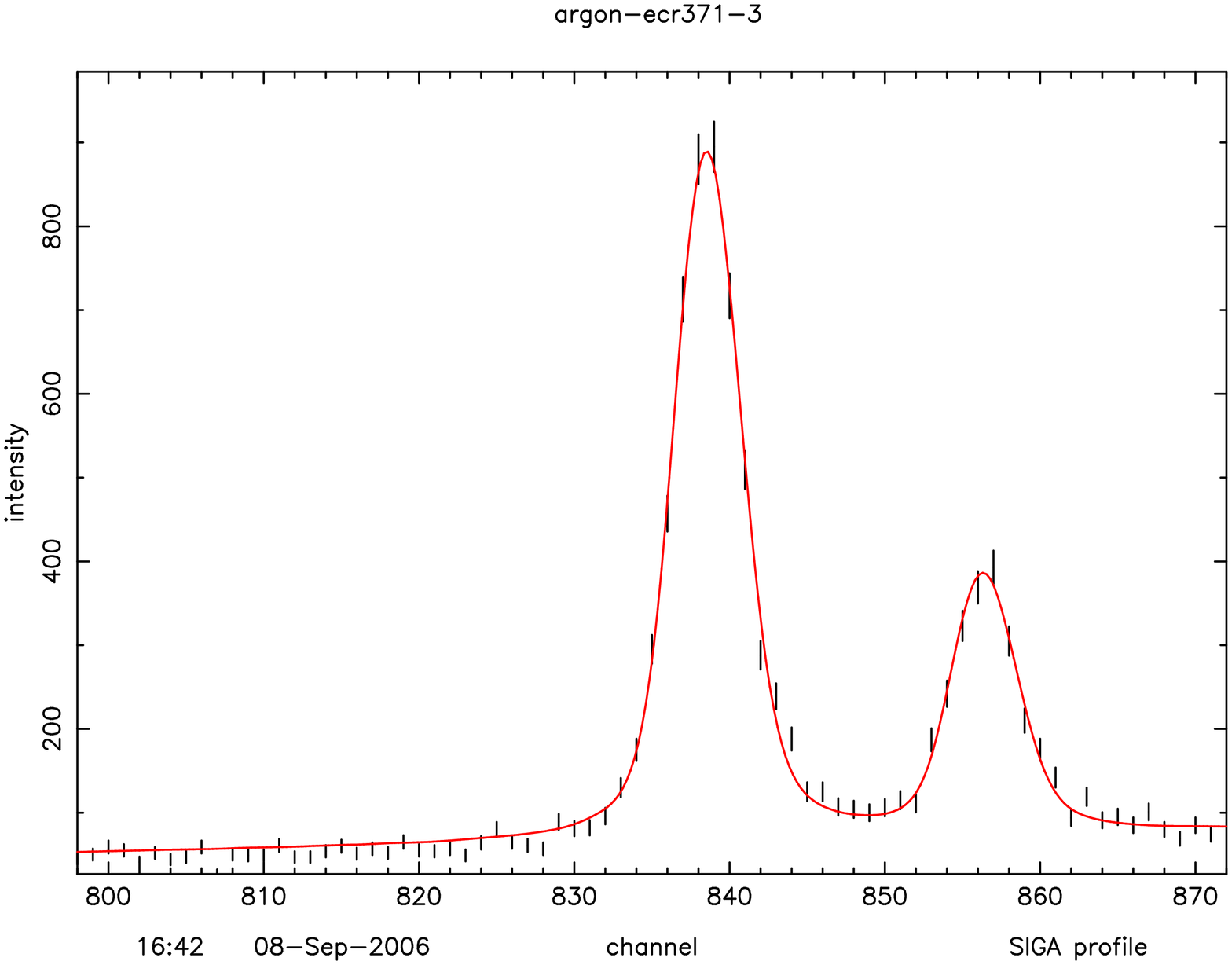}
\includegraphics[width=4.5cm,angle=270]{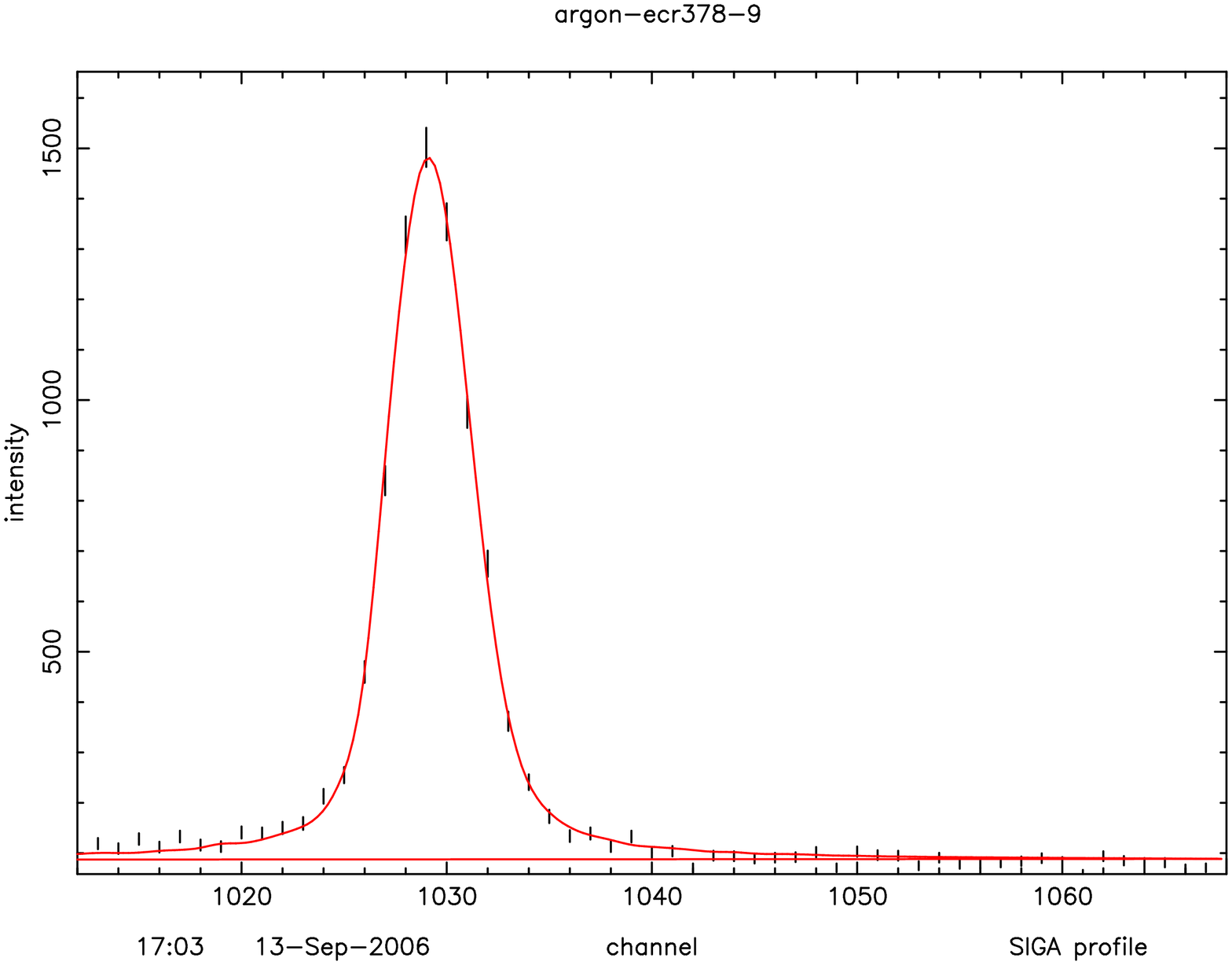}
\caption[]{ 
The $1s 2s 2p \,^2P_{J} \to 1s^2 2s \,^2S_{1/2}$, $J=1/2$, $3/2$ doublet in lithium-like Ar (top) and the relativistic M1 transition ($1s 2s \,^3S_1\to 1s^2\,^1S_0$ in helium-like
Ar (bottom, fitted with the profile discussed in the text.
}
\label{fig:li-like-ar}
\end{figure}

With  the ECR source a number of 20000 events was reached for the 
narrow M1 transition of helium-like Argon 
 in about 30   minutes time to be compared with a number 
of 5000 counts reached after  40 hours  with X-rays from pionic 
 carbon formed when using methane gas. A  total of  about 10 hours was needed, however, 
to determine the  spectrometer's response function  in sufficient 
detail  including changes of the distance CCD detector-crystal (focal scans) and changing 
 apertures in front of the crystals. A complete survey of all K$\alpha$ transitions for ions
from helium-like to $1+$ was performed \citep{tbbc2005}. In the present paper, we will foccus
on a preliminary measurement of the energy difference between the lithium-like $1s 2s 2p\,^2P_J \to 1s^2 2s\,^2S_{1/2}$
and the helium-like M1 transition, used as a reference.  Preliminary results for He-like lines have been presented
elsewhere \citep{tbbc2005,tbcg2006}. For the energy value
of this line, we take recent results from Ref.~\citep{asyp2005}, which includes all QED corrections
known to date. A complete Monte-Carlo simulation of the spectrometer is performed, in order
to provide the line profiles to be fitted \citep{abgg2005,tbcg2006}. An example of spectra,
together with the fit, is presented on Fig.~\ref{fig:li-like-ar}.

In order to interpret the results, which are of much better accuracy than all previous measurements, 
we have performed Multi-Configuration Dirac-Fock calculations. In this variational method
the wave-function is represented as a linear combination of Slater determinants (configurations).
We use the program of Desclaux and Indelicato \citep{des1993,ind1995} that contains first and second 
order QED corrections, including self-energy screening. This code was used in an earlier
calculation of all argon lines \citep{cmps2001,mcsi2001}  needed to interpret spectra from an ECRIS \citep{dkgb2000}.
Here we have expanded on this previous work, by adding all the configurations that 
can be generated from all single and double excitations up to the $n=3$ shell. In one case
we performed a calculation including all correlation up to the $n=4$ shell. From that,
we can deduce that our calculation is accurate to within 40~meV. The self-energy screening has
been evaluated by two different methods, one based on the Welton approximation, and the 
other one based on the direct evaluation of the QED diagrams\citep{iam2001}. This show that we can expect
an uncertainty around 40~meV from QED.
 The results of the calculation for one transition in Ar, 
and a comparison with the experimental value are presented on Table \ref{tab:calc}.

\begin{table*}[t]
\caption{ 
Theoretical contributions to the $1s 2s 2p \,^2P_{1/2}\to 1s^2 2s \,^2S_{1/2}$ in Li-like Ar (eV).
Numbers in parenthesis represent uncertainty on the last digits. Coulomb: Dirac-Fock Coulomb contribution.
Magnetic, Retardation: Dirac-Fock contribution from the magnetic and retarded part of the relativistic electron-electron interaction in the Breit
approximation. Higher-order ret.: retardation corrections beyond Breit interaction. S.E.: one-loop self-energy. Screen (Welton): self-energy screening
correction in the Welton  approximation \citep{igd87,iad90}. V11: Vacuum polarization in the Uehling 
approximation (see e.g., \citep{far1976,bai2000}), 
V13 Wichmann and Kroll correction \citep{wak56,far1976}.  2nd order QED: sum of all two-loop radiative corrections.
}\label{tab:calc}                       

\renewcommand{\arraystretch}{0.9} 
\begin{tabular}{lf{3}f{3}f{7}}
\hline
Contribution  &  \multicolumn{1}{c}{$1s 2s 2p \,^2P_{1/2}$}  &  \multicolumn{1}{c}{$1s^2 2s \,^2S_{1/2}$}  &  \multicolumn{1}{c}{transition}    \\
\hline
Coulomb  &  -6353.4545  &  -9468.1885  &  3114.734    \\
Magnetic  &  0.1800  &  2.2029  &  -2.023    \\
Retardation  &  0.0479  &  -0.0216  &  0.069    \\
Higher-order ret.  &  0.0012  &  0.0002  &  0.001    \\
Coul. + Breit Corr.  &  -0.6318  &  -1.4534  &  0.822  (41)  \\
S.E.  &  1.3797  &  2.5972  &  -1.218    \\
Screen (Welton)  &  -0.0381  &  -0.1226  &  0.085    \\
V11  &  -0.0938  &  -0.1732  &  0.079    \\
V13  &  0.0003  &  0.0006  &  0.000    \\
2nd order QED  &  -0.0012  &  -0.0021  &  0.001    \\
Recoil  &  0.0001  &  0.0002  &  0.000    \\
\hline
Total  &  -6352.6103  &  -9465.1604  &  3112.550    \\
Experiment  &    &    &  3112.453  (8)  \\
Obs.-Calc.  &    &    &  -0.097    (8) (41) \\
\hline
SE Screen (Ref. \citep{iam2001})  &  -0.0313  &  -0.1530  &  0.122    \\
Total  &  -6352.6035  &  -9465.1908  &  3112.587    \\
Obs.-Calc.  &    &    &  -0.134   (8) (41) \\
\hline
\end{tabular}\\[2pt]
\end{table*}

\begin{figure}[t]
\centering
 \includegraphics[scale=0.35,angle=-90,width=\columnwidth]{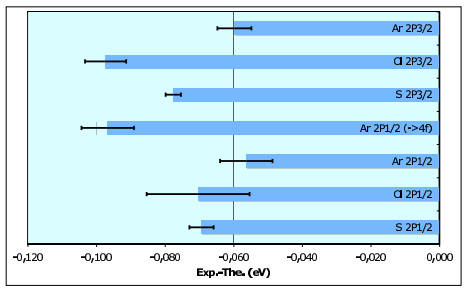}
 \caption[]{
Comparison between the MCDF calculations presented in the text and preliminary experimental results,
for the energy of the $1s 2s 2p \,^2P_J \to 1s^2 2s \,^2S_{1/2}$, $J=1/2$, $3/2$ in lithium-like sulfur, chlorine and argon.
The experimental error bars represent only statistical uncertainty. The average observed-theoretical value difference is 0.072~eV.
Correlation energy is calculated within the $n=3$ active space, except for one case, denoted by $4f$ in which is has been extended to $n=4$.
 }
 \label{fig:exp-th}
 \end{figure}

On Fig.~\ref{fig:exp-th} we present the difference of the
 preliminary experimental transition energies and
the theoretical ones, together with statistical fitting uncertainties. One can see that, even though different crystals have been used, 
and 3 different elements have been measured, this differences are very close.
This show that our results are already rather reliable and our analysis essentially correct.
The average difference of 0.072~eV can be explained by missing correlation contribution, due to the limited basis set used,
by the limitation of the QED corrections. One should also remember that the Li-like transitions that we are studying start from 
auto-ionizing levels. There should then be an Auger broadening, much smaller indeed than for the neutral case, because of the
very small number of allowed decay channels. But there must be an Auger shift, of the same order of magnitude.
The $1s 2s 2p \,^2P_{3/2} \to 1s^2 2s \,^2S_{1/2}$ has a radiative width of 66~meV and an Auger width of  6~meV, while for the 
$1s 2s 2p \,^2P_{1/2} \to 1s^2 2s \,^2S_{1/2}$ these values are 57~meV and 65~meV, respectively \citep{cmps2001}.
To our knowledge Auger shifts have been calculated only for neutral atoms with a K, L or M hole \citep{ibl1998,dkib2003}.

\section{Conclusion and Outlook}

We have presented preliminary values for the transition energies 
of Li-like of transition energy for Li-like ions, using the relativistic M1 transition from He-like ion as a reference.
We get very accurate results, that compare quite well with theory. These measurements demonstrate the potential of the newly-developed ECRIT
at PSI for studies of highly charged ions. They provide very interesting testing ground for relativistic many-body theory and QED, that
cannot be matched by even the most accurate X-ray measurements on neutral atoms in solid targets.

\section{Acknowledgements}

The suggestions and the help of D. Hitz and K. Stiebing in the 
preparatory phase of the experiment are warmly acknowledged. We also 
thank H. Reist for offering  the 6.4 GHz emitter for free.
Special thanks go to the Carl Zeiss Company in Oberkochen, Germany, which 
manufactured the Bragg crystals.
Laboratoire Kastler Brossel is Unité Mixte de Recherche du CNRS 
n$^{\circ}$ 8552.

\bibliography{refs}

\end{document}